\documentclass{ws-ijmpa}
\usepackage{url}
\usepackage[super,compress]{cite}
\usepackage{graphicx}
\usepackage{hyperref}
\hypersetup{
    colorlinks=true,
    linkcolor=blue,
    filecolor=magenta,      
    urlcolor=cyan,
}

\usepackage{epsfig}
\usepackage{epstopdf}
\usepackage{color}

\newcommand{\vslash}[1]{#1 \hspace{-0.5 em} /}

\begin{document}
\markboth{I.V. Truten \& A.Yu. Korchin}{Energy and angular distributions of the bottom quark}

%
\catchline{}{}{}{}{}
%

\title{Energy and angular distributions of the bottom quark in the electron--positron annihilation $e^+e^-\to b \, W^+ \, \bar{t}$}

\author{I.V.~Truten $^{1, *}$\, ,  
A.Yu.~Korchin $^{1,2,\dag}$}

\address{$^1 $NSC ``Kharkov Institute of Physics and Technology'',  61108 Kharkiv, Ukraine\\
$^2$ V.N.~Karazin Kharkiv National University,  61022 Kharkiv, Ukraine\\
\href{mailto:i.truten@kipt.kharkov.ua}{\footnotemark{} i.truten@kipt.kharkov.ua}\\
\href{mailto:korchin@kipt.kharkov.ua}{\footnote{Corresponding author} korchin@kipt.kharkov.ua}
}

\maketitle

\begin{history}
\received{10 September 2020}
\revised{24 October 2020}
\accepted{24 November 2020}
\published{26 January 2021}
\end{history}

\begin{abstract}
The distributions of the bottom quark in the process $e^+ e^- \to t \, \bar{t} \to  b \, W^+ \, \bar{t}$
are considered at the $e^+ e^-$ energy corresponding to the first construction stage of the Compact Linear Collider.  
The cross-sections of $e^+ e^- \to t \, \bar{t} \to  b \, W^+ \, \bar{t}$, as functions of the $b$-quark energy and angle 
with respect to the direction of the electron beam, are derived and calculated. 
The effects of physics beyond  the Standard Model are included via the modified $\gamma t \bar{t}$ and $Z t \bar{t}$ couplings  
which naturally appear in effective field theories.  In addition to the cross-sections, the energy and angular asymmetries   
are calculated. The dependence of these observables on the $e^+ e^-$ energy is calculated, and features of this dependence 
are investigated.

\end{abstract}

\ccode{PACS numbers: 12.15.-y, 12.60.-i, 14.65.Fy, 14.65.Ha}

\section{\label{sec:Introduction}Introduction}

Nowadays the global interest in particle physics is the search for ``new physics'', or  
physics beyond the Standard Model (SM). An important direction of research is study of properties of the top quark.  
These properties are planned to be explored precisely on future electron--positron colliders, such as International 
Linear Collider (ILC) \cite{Behnke:2013xla} and Compact Linear Collider (CLIC) \cite{Aicheler:2012bya, Zarnecki:2019vrn, Zarnecki:2020ics}. 
The ILC will start at the center-of-mass (CM) energy of 250 GeV followed by 500 GeV upgrade \cite{Zarnecki:2020ics, Evans:2017rvt}. 
The CLIC promises to be a good candidate for production of the on-mass-shell top quark and studying its properties.  
At the first construction stage of the CLIC,  the CM energy will be 380 GeV with expected integrated luminosity of 1 ab$^{-1}$, 
which will include 100 fb$^{-1}$ collected near the $t\bar{t}$ production threshold \cite{Zarnecki:2019vrn, Zarnecki:2020ics, CLIC:2016zwp}.

This work is a continuation of our previous paper \cite{Truten:2019sqb}, where the polarization of the top quark, 
produced in electron--positron annihilation,  was studied in detail with emphasis on physics beyond the SM (BSM).  
The aim of the present paper is to consider the consequent decay of the top quark 
$e^+ e^- \to t \, \bar{t} \to  b \, W^+ \, \bar{t}$. Clearly, the observables related to the bottom quark are 
more appropriate for the future experimental investigations. 

Note that this reaction was studied in Refs.~\citen{Arens:1994jp, Arens:1992wh, Grzadkowski:1996kn, Grzadkowski:2001tq}, 
where the distributions of the lepton, coming from the decay $W^+ \to \ell^+ \nu_{\ell}$, were evaluated. 
The issue of a possible $CP$ violation was studied. $CP$ violation in framework of the MSSM was discussed 
and estimated in the spectra of the bottom quark in Refs.~\citen{Christova:1998et,Bartl:1995gj,Bartl:1998ja, Bartl:1998nn}.  

In this paper, we concentrate on distributions of the bottom quark. In addition to the SM, we take into account the    
BSM effects which are described by the anomalous interactions of the photon and $Z$ boson with the top quark. 
These interactions naturally appear in effective field theory (EFT) Lagrangian (see, e.g., Ref.~\citen{AguilarSaavedra:2008zc}),
which contains both the SM Lagrangian and the higher-dimensional terms beyond the SM.    
The calculations in our paper are performed using the formalism developed in Ref.~\citen{Kawasaki:1973hf} which allows one to 
find in a compact form distributions of the secondary particles, such as the bottom quark in $e^+  e^- \to t \, \bar{t} \to b \,  W^+ \, \bar{t}$.   

We investigate effects of the BSM couplings $\kappa$ and $\kappa_z$, which determine the anomalous $\gamma t \bar{t}$ 
and $Z t \bar{t}$ vertices,  and influence of the polarization of the top quark arising  
in the $e^+ e^- \to t \, \bar{t}$ process on the energy and angular distributions of the $b$ quark.  
In this connection note, that in framework of the SM, the distribution of the bottom quark was analytically derived in Ref.~\citen{Arens:1992wh} 
for the unpolarized leptons and in Ref.~\citen{Christova:1997by} for the longitudinally polarized leptons. 
In this work we perform calculation beyond the SM, where analytical consideration is rather cumbersome, and using the 
formalism~\cite{Kawasaki:1973hf} facilitates calculations.   
   
The structure of the paper is as follows.  In Section \ref{sec:formalism} theoretical basis for calculation of the 
process $e^+ e^- \to t \, \bar{t} \to  b \, W^+ \, \bar{t}$ is overviewed.  Two possible ways of calculation of the cross-section suggested 
in Ref.~\citen{Kawasaki:1973hf} are mentioned. We also introduce polarization vector of the top quark which arises 
in the production process $e^+ e^- \to t \, \bar{t}$ \cite{Truten:2019sqb}.
In Subsection \ref{subsec:energy} the cross-section of $e^+ e^- \to  b \, W^+ \, \bar{t}$ is considered as a function of the bottom-quark energy $E_b$,  
while in Subsection \ref{subsec:angular} the cross-section is obtained as a function of the polar angle $\theta_b$ between the momentum of 
$b$ quark and direction of electron beam. 
In Section \ref{sec:results} results of calculation in the SM and BSM of the $e^+ e^- \to  b \, W^+ \, \bar{t}$ cross-sections as functions of the energy $E_b$ 
and the angle $\theta_b$ are presented.  The BSM results are  obtained for some values of the coupling constants $\kappa$ and $\kappa_z$.   
Several observables, such as normalized energy and angle distributions, and energy and angular asymmetries are  
studied at various values of $\kappa$ and $\kappa_z$. Dependence of asymmetries on the $e^+ e^-$ invariant energy is investigated.    
In Section \ref{sec:conclusions} conclusions are given. \ref{app:A} contains a proof of the gauge invariance of the photon-exchange diagram for the process $e^+ e^- \to t \, \bar{t} \to  b \, W^+ \, \bar{t}$ in the narrow-width approximation.


\section{\label{sec:formalism} Formalism for the Process $e^+ \, e^- \to  b \, W^+ \, \bar{t}$ }

Let us consider electron--positron annihilation into a pair of top quarks, where one of the quarks decays, $t \to b \, W^+$, {\it i.e.} 
the process $e^+  e^- \to t  \, \bar{t} \to b \, W^+ \, \bar{t}$.
To calculate the energy and the angular distribution of the $b$ quark one can use efficient formalism of Refs.~\citen{Kawasaki:1973hf, Arens:1994jp}. 
According to these papers there are two equivalent ways of calculation of the 
cross-section of interest.  In the first way the cross-section is written as 
\begin{equation}
 d \sigma (e^+e^- \to   b \, W^+ \, \bar{t}) = \int  \frac{d \sigma (e^+ e^- \to t \, \bar{t}; \, 0)}{d \Omega_t}  \,
\frac{d \Gamma ({t \to b \, W^+}; \, a^\mu)}{\Gamma_t} \, d \Omega_t,
\label{eq:001}
\end{equation}
where ${d \sigma (e^+ e^- \to t \,  \bar{t}; \, 0)}/{d \Omega_t} $ is the differential cross-section of the electron--positron annihilation to the top quarks, which is 
calculated for all unpolarized particles. $d \Gamma ({t \to b \, W^+}; \, a^\mu)$ is the differential width of the decay of the polarized top quark, 
where its polarization arises in the process $e^+ e^- \to t \, \bar{t}$ and is described by the four-vector $a^\mu$.  
The cross-section is calculated in the CM frame, while the  $t \to b \, W^+$ differential decay width and the total width 
$\Gamma_t$ are evaluated in the top-quark rest frame.      

The second way of calculation implies that the cross-section can be presented as
\begin{equation}
 d \sigma (e^+e^- \to   b \, W^+ \, \bar{t}) = 2 \int  \frac{d \sigma (e^+ e^- \to t \, \bar{t}; \, n^\mu)}{d \Omega_t}  \,
\frac{d \Gamma ({t \to b \, W^+}; \, 0)}{\Gamma_t} \, d \Omega_t,
\label{eq:002}
\end{equation}
where the $e^+ e^- \to t \, \bar{t}$ \ differential cross-section is calculated for the polarized top quark with the polarization 
four-vector $n^\mu$ 
\begin{equation}
n^\mu = \alpha_b  \left( \frac{p_b^\mu m_t}{p_t \cdot p_b} - \frac{p_t^\mu  }{m_t} \right),   \quad \qquad n \cdot p_t =0,  \qquad n \cdot n = - {\alpha_b}^2, 
\label{eq:003}
\end{equation}
where $p_t^\mu (p_b^\mu)$ is the four-momentum of the top (bottom) quark, and the parameter $\alpha_b$ determines the asymmetry 
in the decay $t \to b \, W^+$ of the polarized top quark.  Also in Eq.~(\ref{eq:002}) 
$d \Gamma( {t \to b \, W^+}; \, 0) = \frac{1}{2} \sum_s d\Gamma^{(s)}$ is the initial spin-averaged differential decay width.

Below we will mainly use the form in Eq.~(\ref{eq:001}), since only the $e^+ e^- \to t \, \bar{t}$ unpolarized cross-section enters, 
while the polarization vector $a^\mu$ of the produced top quark was already calculated in Ref.~\citen{Truten:2019sqb}. 
In Sec.~\ref{sec:results} we check explicitly the equivalence of the cross-sections in Eqs.~(\ref{eq:001}) and (\ref{eq:002}).  

Let us discuss the distribution of the bottom quark. In general, the differential decay width of $t \to b \, W^+$ is
\begin{equation}
 d \Gamma = \frac{1}{2 m_t}|{\cal M}|^2 \, dX_{LIPS}
\label{eq:004}
\end{equation}
with $dX_{LIPS}$ being the Lorentz invariant phase space, and the matrix element squared for the polarized top  quark and the unpolarized $W$ boson and $b$ quark  reads
\begin{equation}
|{\cal M}|^2 = \sqrt{2} \, G_F  N \, |V_{tb}|^2 (1 + \alpha_b \, \vec{P} \vec{n}_{b, R}) =  
\sqrt{2} \, G_F  N |V_{tb}|^2 \left(1 - \alpha_b \, \frac{ m_t \, a \cdot p_b}{p_t \cdot p_b} \right),    
\label{eq:005}
\end{equation}
where $G_F$ is the Fermi weak constant, $V_{tb} \approx 1 $ is the element  of the CKM matrix,  and
\begin{equation}
N =  (m_t^2-m_W^2) (2m_W^2 +m_t^2), \qquad \quad \alpha_b = \frac{2m_W^2 -m_t^2}{2m_W^2 +m_t^2} =  - 0.40. 
\label{eq:006}
\end{equation}
Here $m_t$ ($m_W$) is the mass of the top quark ($W$ boson) and we neglect the bottom-quark mass, {\it i.e.} put $m_b=0$. 
Note that Eq.~(\ref{eq:005}) is written both in the rest frame of the top quark (denoted by the subscript ``R'') and in the Lorentz-invariant form.  
In the frame ``R'',  \ $\vec{n}_{b, R}$ is the unit vector in the direction of the $b$-quark momentum, and the $t$-quark polarization is determined by the three-vector  
$\vec{P}$, such that $a^\mu_{R}=(0, \, \vec{P})$. In the frame, where $t$ quark moves with the 
momentum $\vec{p}_t$ and energy $E_t$, the polarization four-vector is~\cite{Berestetsky:1982aq}
\begin{equation}
a^\mu= \Bigl( \frac{\vec{P}  \vec{p}_t}{m_t} , \, \vec{P} + \frac{\vec{p}_t \, (\vec{P}  \vec{p}_t) }{m_t \, (m_t + E_t)} \Bigr), \qquad   \quad
a \cdot p_t  =0 , \qquad  a \cdot a = - \vec{P}^{\,2}.
\label{eq:0061}
\end{equation}  

The total decay width is
\begin{equation}
\Gamma_t = \frac{p_b^0 G_F |V_{tb}|^2 N}{4 \sqrt{2} \pi m_t^2},
\label{eq:007}
\end{equation}
 where $p^0_b = ({m_t^2 - m_W^2})/{2 m_t}$  is the bottom-quark momentum (energy) in the $t$-quark rest frame. 
 
The two-particle phase space, 
\begin{equation}
 dX_{LIPS} = (2 \pi)^4 \delta^4 (p_t - p_b - p_W) \frac{d^3 p_b}{(2 \pi)^3 2 E_b} \frac{d^3 p_W}{(2 \pi)^3 2 E_W},
 \label{eq:008}
\end{equation} 
where $E_b$ and $E_W$ is the energy of the bottom quark and $W$ boson, respectively, allows one to integrate over 
the $W$-boson momentum and obtain 
\begin{eqnarray}
  \frac{1}{\Gamma_t} \, d \Gamma (t \to b \, W^+, \, a^\mu) 
&=& \frac{m_t}{4 \pi p_b^0} \left( 1 - \alpha_b \, \frac{m_t \, a \cdot p_b }{p_t \cdot p_b} \right)  \nonumber \\ 
&& \times
\delta(E_t - E_b - E_W) \frac{E_b}{E_W}  dE_b \, d \Omega_b 
\label{eq:009}
\end{eqnarray}
with $E_W = \left( m_W^2 + \vec{p}_t^{\,2} +E_b^2 - 2 \vec{p}_t \, \vec{p}_b \right)^{1/2}$ and $E_b = |\vec{p}_b|$. 
Substitution of (\ref{eq:009}) in (\ref{eq:001}) gives the required cross-section 
\begin{eqnarray}
d \sigma (e^+e^- \to   b \, W^+ \, \bar{t}) &=& \frac{m_t}{4 \pi p_b^0}  \int  \frac{d \sigma (e^+ e^- \to t \bar{t}; \, 0)}{d \Omega_t}  \,
\left( 1 - \alpha_b \, \frac{m_t \, a \cdot p_b }{p_t \cdot p_b} \right) \nonumber \\ 
&& \times  \delta(E_t - E_b - E_W)  \frac{E_b}{E_W} \, dE_b \, d \Omega_b  \, d \Omega_t,
\label{eq:010}
\end{eqnarray}
from which one can obtain the energy and angular distributions of the bottom quark for arbitrary $e^+ e^- \to t \bar{t}$ cross-section.  


\subsection{\label{subsec:energy} Energy spectrum of the bottom quark}

\begin{figure}[tbh]
\centerline{
\includegraphics[width=0.5\textwidth]{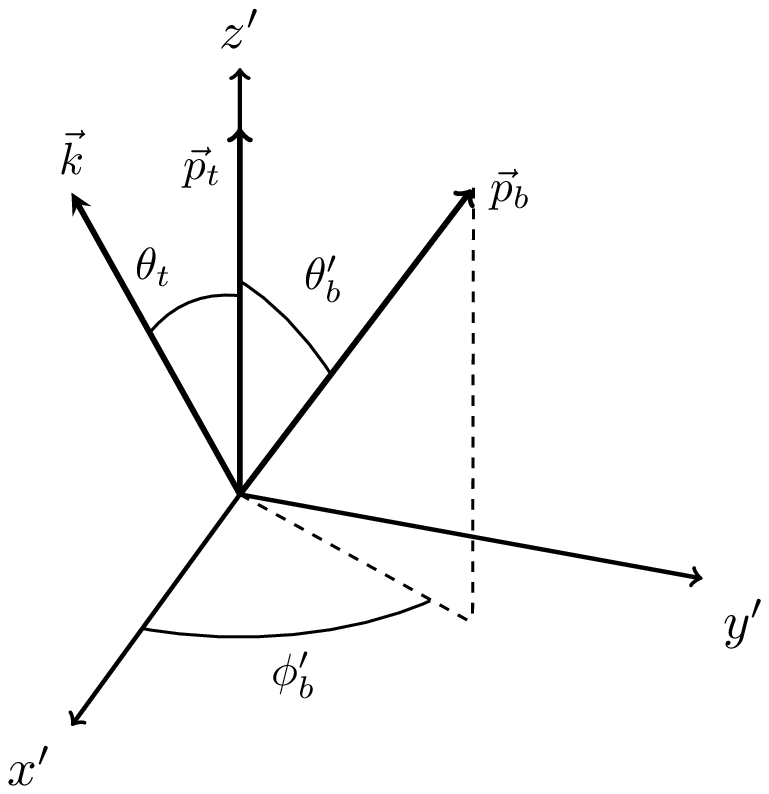}
\includegraphics[width=0.5\textwidth]{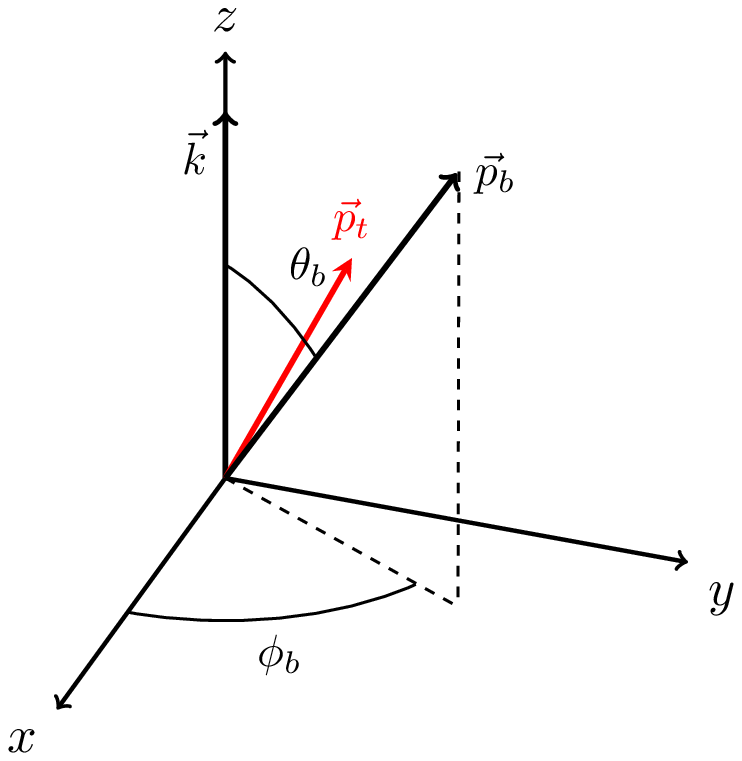}}
\caption{The coordinate systems used for the energy distribution (left) and the angular distribution (right) of the bottom quark.}
\label{fig:2}
\end{figure}

For this case the coordinate system is chosen as shown in Fig.~\ref{fig:2} (left):
the top-quark (and antiquark) momentum is directed along the $OZ^\prime$ axis, the electron (and positron) momentum 
lies in the $X'OZ'$ plane, the bottom-quark momentum points in arbitrary direction. Further, $\theta_t$ is the angle between the electron 
and the top-quark momenta, $\theta_b^\prime$ is the angle between $\vec{p}_t$ and $\vec{p}_b$.

In the CM frame, the four-momenta of $e^-$, $t$ and $b$ quarks, and polarization vector are  
\begin{eqnarray}
k^\mu &=& E_t \bigl(1, \,  \sin{\theta_t}, \, 0, \, \cos{\theta_t} \bigr), \nonumber \\
p_t^\mu &=& E_t \bigl(1,\, 0,\, 0,\, V \bigr), \nonumber \\
p_b^\mu &=& E_b \bigl( 1, \, \sin{\theta^\prime_b} \cos{\phi^\prime_b}, \, \sin{\theta^\prime_b} \sin{\phi^\prime_b}, \,  \cos{\theta^\prime_b} \bigr), \nonumber \\
a^\mu &=& \bigl(\gamma V P_{z^\prime}, \, P_{x^\prime},  \, P_{y^\prime}, \, \gamma P_{z^\prime} \bigr),
\label{eq:011}
\end{eqnarray}
where $V=p_t/E_t$ is the velocity of the top quark, and $\gamma=E_t/m_t$ is the Lorentz factor.  
The components of the polarization vector $\vec{P} = (P_{x^\prime}, \, P_{y^\prime},  \, P_{z^\prime})$  
have been evaluated in Ref.~\citen{Truten:2019sqb} as functions of the top-quark angle $\theta_t$. 
Then one can carry out integration in Eq.~(\ref{eq:010}) over the angles of the $b$ quark $\theta_b^\prime$ and $\phi^\prime_b$, 
and the cross-section as a function of the bottom-quark energy becomes
\begin{eqnarray}
\frac{d \sigma (e^+ e^- \to   b \, W^+ \, \bar{t})}{d E_b} &=& \frac{1 }{2 \gamma V p_b^0} \int_{-1}^{+1}  \frac{d \sigma (e^+ e^- \to t \, \bar{t}; \, 0)}{d \cos \theta_t} \nonumber \\
&&\times 
\Bigl[ 1 + \frac{\alpha_b  \, P_{z^\prime} (\theta_t)}{V} \, \Big( \frac{E_b}{\gamma \, p_b^0} -1 \Big)  \Bigr] \, d \cos \theta_t ,
\label{eq:012}
\end{eqnarray}
where the energy varies within the limits
\begin{equation}
E_{-} \le E_b \le   E_{+},  \qquad \qquad  E_{\pm}=\frac{p_b^0}{\gamma \,  (1 \mp V)}.
 \label{eq:013}
\end{equation}
 It is seen that only the longitudinal component of the quark polarization survives after integration over the azimuthal angle $\phi^\prime_b$.   
The cross-section in Eq.~(\ref{eq:012}) is linear in the energy $E_b$, if we keep the polarization of the top quark. If  
polarization is neglected, {\it i.e.} $ \vec{P}=0$, the cross-section (\ref{eq:012}) is independent of the $b$-quark energy.  
In the SM, this feature was pointed out in Ref.~\citen{Christova:1997by}, and apparently 
this is a general property of the energy spectrum regardless of a model for $e^+ e^- \to t \, \bar{t}$.       
It is also interesting that the $t$-quark polarization does not contribute at the $b$-quark energy $\tilde{E}_b = \gamma \, p_b^0$, 
as is seen from the last term in (\ref{eq:012}).    

Using Eq.~(\ref{eq:012}) we can calculate the total cross-section by integrating (\ref{eq:012}) over  $E_b$. 
Then use of (\ref{eq:013}) yields  
\begin{eqnarray}
\int_{E_-}^{E_+} &&\frac{d \sigma (e^+ e^- \to   b \, W^+ \, \bar{t})}{d E_b} \, dE_b \nonumber \\
&& = \int_{-1}^{+1}  \frac{d \sigma (e^+ e^- \to t \, \bar{t}; \, 0)}{d \cos \theta_t} \, d \cos \theta_t  = \sigma (e^+ e^- \to t \, \bar{t}), 
 \label{eq:014}
\end{eqnarray}
which is the normalization of the cross-section.


\subsection{\label{subsec:angular} Angular spectrum of the bottom quark}

It is convenient to define the angular distribution of the bottom quark in the coordinate system shown in Fig.~\ref{fig:2} (right).   
In this system the polar angle of the $b$ quark is determined with respect to the direction of the electron beam which is parallel to the $OZ$ axis. 
The $b$-quark momentum is defined by the polar angle $\theta_b$ and the azimuthal angle $\phi_b$.

The system in Fig.~\ref{fig:2} (right) is obtained from the system in Fig.~\ref{fig:2} (left) by the clockwise rotation around the $OY^\prime$ axis on  
the angle $\theta_t$, so that  
\begin{equation}
\left( 
\begin{array}{cccc}
  x\\
  y\\
  z
 \end{array}
 \right)
 = \left(
\begin{array}{cccc}
\cos{\theta_t} & 0 & -\sin{\theta_t}\\
0 & 1 & 0\\
\sin{\theta_t} & 0 & \cos{\theta_t}\\
\end{array}
\right)
\left( \begin{array}{cccc}
  x^{\prime}\\
  y^{\prime}\\
  z^{\prime}
 \end{array}\right),
\label{eq:015}
\end{equation}
where $x^\prime, y^\prime, z^\prime$ are the primary non-rotated axes and $x, y, z$ are the rotated ones.
The four-momenta of the particles and the polarization vector take the form 
\begin{eqnarray}
k^\mu &=& E_t \bigl(1, \, 0,\, 0,\, 1 \bigr), \nonumber \\
p_t^\mu  &=& E_t \bigl(1, \, -V  \sin{\theta_t},\, 0,\, V  \cos{\theta_t} \bigr), \nonumber \\
p_b^\mu  &=& E_b \bigl(1, \,  \sin{\theta_b} \cos{\phi_b}, \,  \sin{\theta_b} \sin{\phi_b}, \,  \cos{\theta_b} \bigr), \nonumber \\
a^\mu &=& \bigl(\gamma V P_{z^\prime}, \, P_{x^\prime} \cos{\theta_t}  - \gamma  P_{z^\prime} \sin{\theta_t} ,  \, P_{y^\prime}, \, P_{x^\prime} \sin{\theta_t}  
+ \gamma P_{z^\prime}  \cos{\theta_t} \bigr),
\label{eq:016}
\end{eqnarray}
 and for the scalar product $a \cdot p_b$ in (\ref{eq:010}) we find
\begin{eqnarray}
a \cdot p_b &=& - E_b \, \left[
P_{x^\prime} \, (\cos \theta_t \sin \theta_b \cos \phi_b  + \sin \theta_t \cos \theta_b) \right. \nonumber \\
&& \left. + P_{y^\prime} \, \sin \theta_b \sin \phi_b - P_{z^\prime} \, \gamma \, (V  - \cos \Theta_{\vec{p}_t \vec{p}_b}) \right],  \label{eq:017} \label{eq:018} \\ 
\cos \Theta_{\vec{p}_t \vec{p}_b} &\equiv&   \cos \theta_t \cos \theta_b - \sin \theta_t \sin \theta_b \cos \phi_b. \nonumber 
\end{eqnarray}

Performing integration in (\ref{eq:010}) over the energy $E_b$ and the azimuthal angle $\phi_b$, we obtain the cross-section as a function of the 
polar angle of the bottom quark
\begin{eqnarray}
\frac{d \sigma (e^+e^- \to   b \, W^+ \, \bar{t})}{d \cos \theta_b} \nonumber \\ 
&=& \int_{-1}^{+1}  d \cos \theta_t \,
\frac{d \sigma (e^+ e^- \to t \, \bar{t}; \, 0)}{d \cos \theta_t}  \int_{0}^{2 \pi} 
 \frac{d \phi_b }{(1-V \cos \Theta_{\vec{p}_t \vec{p}_b})^2}   \nonumber \\
&& \times \frac{1}{4 \pi \gamma^2} \left[1 +  \alpha_b \, \frac{  P_{x^\prime} (\theta_t) \, (\cos \theta_t \sin \theta_b \cos \phi_b  
+ \sin \theta_t \cos \theta_b)}{\gamma \, (1-V \cos \Theta_{\vec{p}_t \vec{p}_b})} \right. \nonumber \\
&& \left.  + \alpha_b \, \frac{P_{y^\prime} (\theta_t) \, \sin \theta_b \sin \phi_b 
- P_{z^\prime} (\theta_t) \, \gamma \, (V  - \cos \Theta_{\vec{p}_t \vec{p}_b})}{\gamma \, (1-V \cos \Theta_{\vec{p}_t \vec{p}_b})}\right].
\label{eq:019}
\end{eqnarray}   
The normalization of this cross-section is
\begin{equation}
\int_{-1}^{+1} \frac{d \sigma (e^+e^- \to  b \, W^+ \, \bar{t})}{d \cos \theta_b} d \cos \theta_b =  \sigma (e^+ e^- \to t \, \bar{t}; \, 0).
\label{eq:020}
\end{equation}


\section{\label{sec:results} Results of calculation and discussion}

\subsection{\label{subsec:cross-sections} cross-sections}

\begin{figure}[tbh]
\centerline{\includegraphics[width=\textwidth]{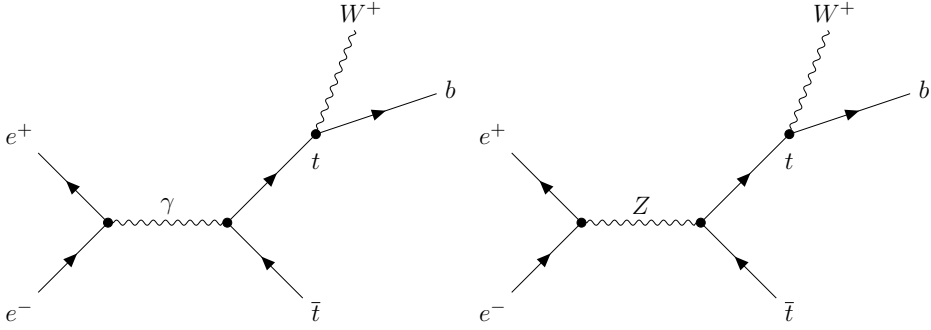}}
\caption{Feynman diagrams for $e^+ \, e^- \to  t \bar{t} \to  b  \, W^+  \bar{t}$ reaction. 
The vertices  $\gamma t \bar{t}$ and $Z t \bar{t}$ can include RC  and BSM couplings.}
\label{fig:1}
\end{figure}

The considered process on the tree level is described by the Feynman diagrams in Fig.~\ref{fig:1}. 
We start with description of this process in the framework of the SM without radiative corrections (RCs) to the vertices 
$\gamma t \bar{t}$ and $Z t \bar{t}$, that corresponds to $\kappa=\kappa_z=0$.  
Then we add RC in the SM by choosing nonzero values of $\kappa$ and $\kappa_z$, and finally include  
anomalous couplings of quarks with the photon and $Z$ boson related to the BSM physics. 
Thus the structure of the $\gamma t \bar{t}$ and $Z t \bar{t}$ vertices is chosen in the form 
\begin{eqnarray}
\Gamma_{\gamma t \bar{t}}^\mu &=& -i e  
\Big[ Q_t \gamma^\mu + i \frac{ \sigma^{\mu \nu} q_\nu}{2 m_t} (\kappa + i \tilde{\kappa}  \gamma_5) \Big] \equiv -ie V_{\gamma t \bar{t}}^\mu, 
\label{eq:020_vertex_g} \\ 
\Gamma_{Z t \bar{t}}^\mu &=& -i \frac{g}{2 \cos \theta_W} \Big[ \gamma^\mu (v_t-a_t \gamma_5)  + i  \frac{\sigma^{\mu \nu} q_\nu}{2 m_t} 
(\kappa_z + i \tilde{\kappa}_z  \gamma_5) \Big]  \nonumber \\
&\equiv& -i \frac{g}{2 \cos \theta_W} V_{Z t \bar{t}}^\mu  ,  
\label{eq:020_vertex_Z}
\end{eqnarray}
where $e$ is the positron charge,  $g = e / \sin \theta_W$ with $\theta_W$ denoting the weak mixing angle, $Q_t =2/3$,   
$v_t= 1/2 -4/3 \sin^2 \theta_W$, $a_t=1/2$,  and $q^\nu= k^\nu + k^{\prime \nu}$ is the four-momentum 
of the intermediate photon ($Z$ boson). 
In the following we neglect the terms in (\ref{eq:020_vertex_g}) and  (\ref{eq:020_vertex_Z}) proportional to $\tilde{\kappa}$ and $\tilde{\kappa}_z$   
responsible for the $CP$ violation, and keep only the couplings $\kappa$ and $\kappa_z$ related respectively to the anomalous magnetic 
and anomalous weak-magnetic dipole moments of the $t$ quark. For details on relation of these  
couplings to the Wilson coefficients in the EFT Lagrangian and some constraints on their values see Ref.~\citen{Truten:2019sqb}.   

RC in the SM to the $\gamma t \bar{t}$ and $Z t \bar{t}$ vertices are contained in Eqs.~(\ref{eq:020_vertex_g}), (\ref{eq:020_vertex_Z})  if 
one chooses the corresponding values, $\kappa_{RC}$ and $\kappa_{z, \, RC}$.  
In this work we do not take into account other RC to the $e^+ e^- \to t \, \bar{t}$ reaction,  
as well as to the $t \to b \, W^+$ decay.  For $e^+ e^- \to t \, \bar{t}$, certain RC have been studied   
in Refs.~\citen{Korner:1993dy, Groote:1995yc, Groote:1995ky, Stav:1996ep, Olsen:1997sk}.   

In general, the diagrams in Fig.~\ref{fig:1} do not satisfy the condition of gauge invariance, and additional tree-level diagrams 
are needed to satisfy the gauge invariance \cite{Boos:1999qc, Boos:2001sj, Boos:2012vm}.  In the approach
used in the present paper, it is assumed following~\cite{Kawasaki:1973hf, Arens:1994jp} that the intermediate $t$ quark, 
decaying to $b$ quark and $W$ boson, is on its mass shell. 
This assumption is realized via the narrow-width approximation (NWA) which is used in derivation of 
Eqs.~(\ref{eq:001}) and (\ref{eq:002}).  In \ref{app:A} we show explicitly that in this approach  
the photon-exchange diagram in Fig.~\ref{fig:1} satisfies the gauge invariance.       

\begin{figure}[tbh]
\centerline{\includegraphics[width=0.70\textwidth]{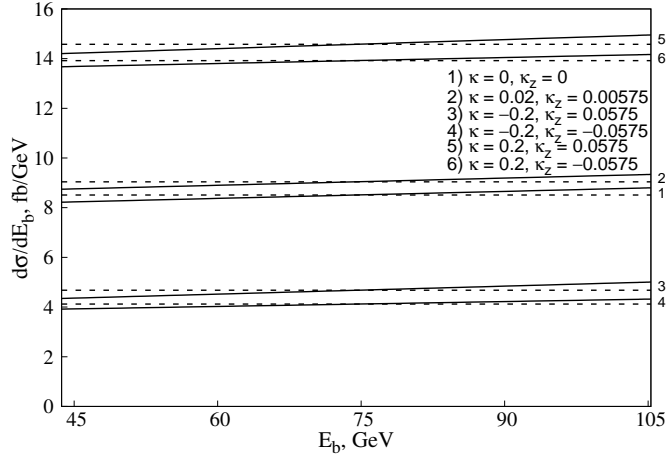} }
\caption{The cross-section of the process $e^+ \, e^- \to   b \, W^+ \, \bar{t}$ as a function of the bottom-quark energy: 
(1) in the SM without RC, (2) in the SM with RC, and (3)-(6) beyond the SM. The 
$e^+ e^-$ energy is $\sqrt{s}=380$ GeV.  Solid (dashed) curves correspond to the polarized (depolarized) process.}
\label{fig:3}
\end{figure}

The calculated cross-section (\ref{eq:012}) is shown in Fig.~\ref{fig:3} for the $e^+ e^-$ energy 380 GeV.   
The energy of the bottom quark lies in the interval 43.7 GeV $ \le E_b \le $ 105.3 GeV.  
We present two variants of the calculation: the first one corresponds  to neglect of the $t$-quark 
polarization, $\vec{P} = 0$ (called ``depolarized'' process), and 
the second one corresponds to inclusion of the polarization, $\vec{P} \neq 0$  (called ``polarized'' process). 
The corresponding curves in Fig.~\ref{fig:3} (dashed and solid) cross at the 
energy $\tilde{E}_b =  \gamma \, p_b^0 = (E_+ + E_-)/2=  74.5$ GeV. 

In Fig.~\ref{fig:3} we also show cross-section for several values of the couplings $\kappa$ and $\kappa_z$.
Note that, firstly, the cross-section varies considerably with the couplings and, secondly, 
the difference between the polarized and depolarized cases is similar to the SM calculation,  
but the slope of the solid straight lines depends on the couplings. This is demonstrated in Table~\ref{tab:1}.

The values of the couplings in Fig.~\ref{fig:3} and Table~\ref{tab:1} are chosen as in Ref.~\citen{Truten:2019sqb}.
Namely, in the first line the values correspond to the SM without RC ($\kappa = \kappa_z =0$), 
while the second line includes RC in the SM in Eqs.~(\ref{eq:020_vertex_g}) and (\ref{eq:020_vertex_Z})
($\kappa_{RC} = 0.02$, \ $\kappa_{z, RC}=0.00575$). The latter were evaluated in Ref.~\citen{Bernreuther:2005gq} 
to the two loops in QCD and to the lowest order in electroweak couplings.
The other values in Fig.~\ref{fig:3} and Table~\ref{tab:1} are chosen ten times bigger than $\kappa_{RC}$ and $\kappa_{z, RC}$, 
as a conservative estimate. Since the sign of the BSM couplings is not known, the negative signs are also included. 
It is seen that the slope of the energy distribution depends on the BSM couplings, although the dependence is weak, at least 
for the considered  moderate values of $\kappa$ and $\kappa_z$.   

\begin{table}[tph]
\tbl{The slope of the cross-section $d \sigma (e^+ e^- \to  b \, W^+ \, \bar{t})/d E_b $ for various couplings $\kappa$,  $\kappa_z$. 
The $e^+ e^-$ invariant energy is 380 GeV.}
{\begin{tabular*}{0.47\textwidth}{c@{\extracolsep{\fill}}c c}
 \toprule
$\kappa$ & $\kappa_z$ & slope, \ $10^{-3}$ ${\rm {fb}/{GeV^2}}$ \\
\colrule
0.0 & 0.0 &    $9.4 $ \\
0.02 & 0.00575 &    $9.7 $ \\
0.2 & 0.0575 &  $12.2 $ \\
0.2 & $-$0.0575 &  $8.0 $ \\ 
$-$0.2 & 0.0575 &  $10.7 $ \\ 
$-$0.2 & $-$0.0575 &  $6.6 $ \\ \botrule
\end{tabular*} \label{tab:1}}
\end{table}

\begin{figure}[tbh]
\begin{center}
\includegraphics[scale=0.70]{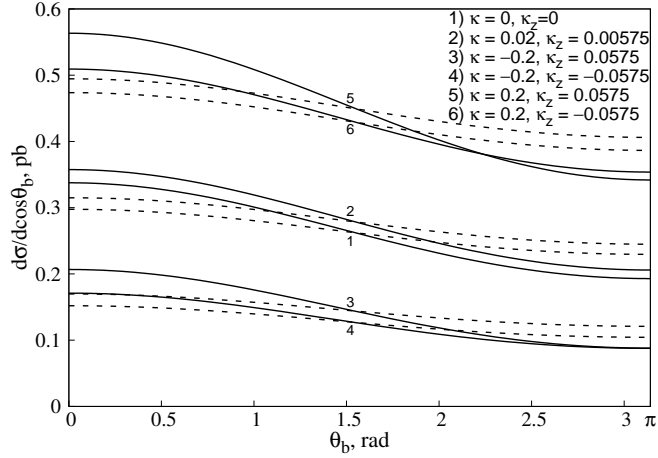}
\end{center}
\caption{The cross-section of the process $e^+ \, e^- \to   b \, W^+ \, \bar{t}$ as a function of the $b$-quark polar angle: (1) in the SM without RC, 
(2) in the SM with RC, and (3)-(6) beyond the SM. Solid (dashed) curves correspond to the polarized (depolarized) process.}
\label{fig:4}
\end{figure}

In Fig.~\ref{fig:4} we show the angular spectrum of the $b$ quark (\ref{eq:019}).   
One can notice that in all calculations the solid and dashed curves cross at the point  
close to the angle $\tilde{\theta} \approx \pi/2$. 
As can also be seen from Fig.~\ref{fig:4},  the difference between the polarized and depolarized processes is more pronounced than 
the corresponding effect in the energy dependence in Fig.~\ref{fig:3}.  Apparently the angular dependence 
is more sensitive to the top-quark polarization. 

It is of interest to plot the energy and angular normalized  distributions 
\begin{equation}
W(E_b) \equiv   \frac{1}{\sigma (e^+ e^- \to t \, \bar{t})}  \frac{d \sigma (e^+ e^- \to   b \, W^+ \, \bar{t})}{d E_b}, 
\label{eq:020_W-energy}
\end{equation}
\begin{equation}
W(\theta_b) \equiv   \frac{1}{\sigma (e^+ e^- \to t \, \bar{t})}  \frac{d \sigma (e^+ e^- \to   b \, W^+ \, \bar{t})}{d \cos \theta_b}, 
\label{eq:020_W-angle}
\end{equation}
which satisfy the normalization
\begin{equation}
\int_{E_-}^{E_+} W(E_b) \, dE_b =  \int_0^\pi W(\theta_b) \sin \theta_b \, d \theta_b =1. 
\label{eq:020_normalization}
\end{equation}

\begin{figure}[tbh]
\centerline{
\includegraphics[width=0.5\textwidth]{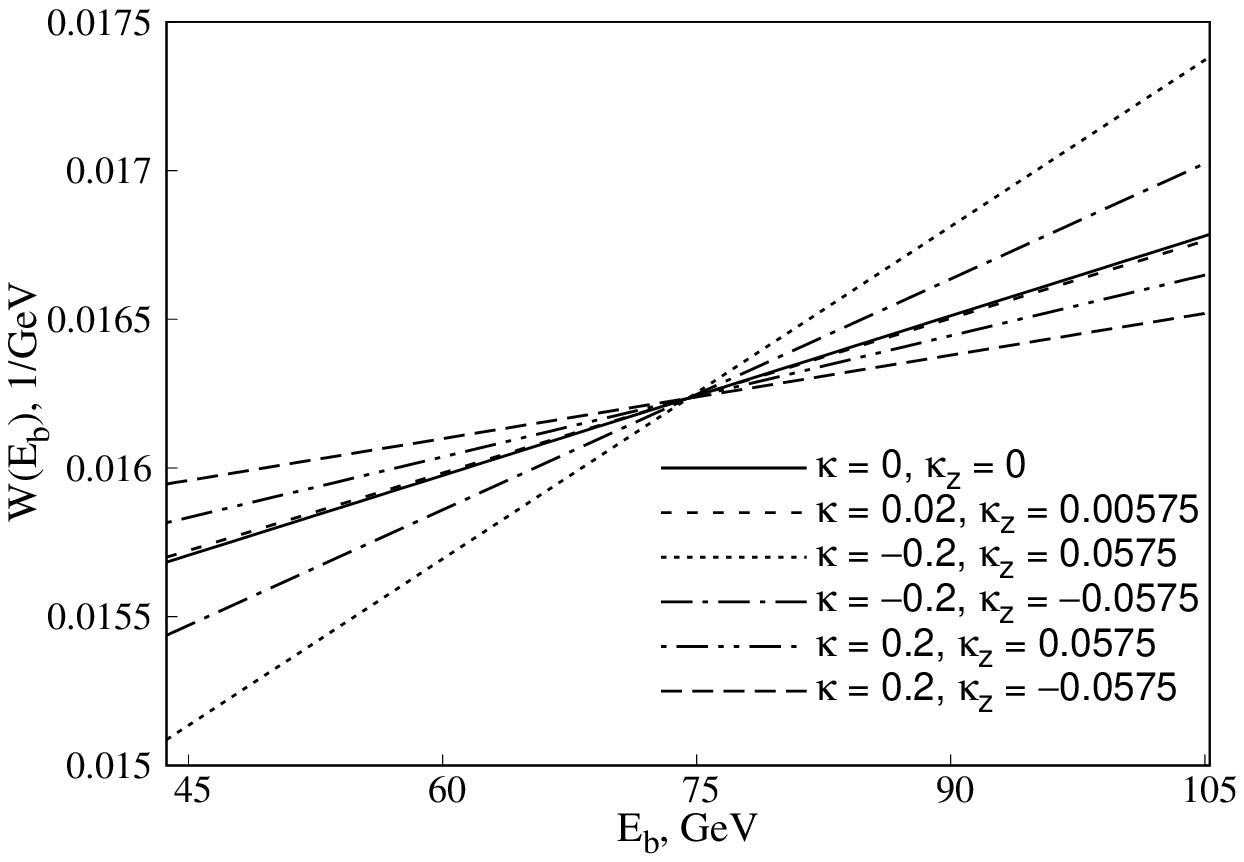}
\includegraphics[width=0.5\textwidth]{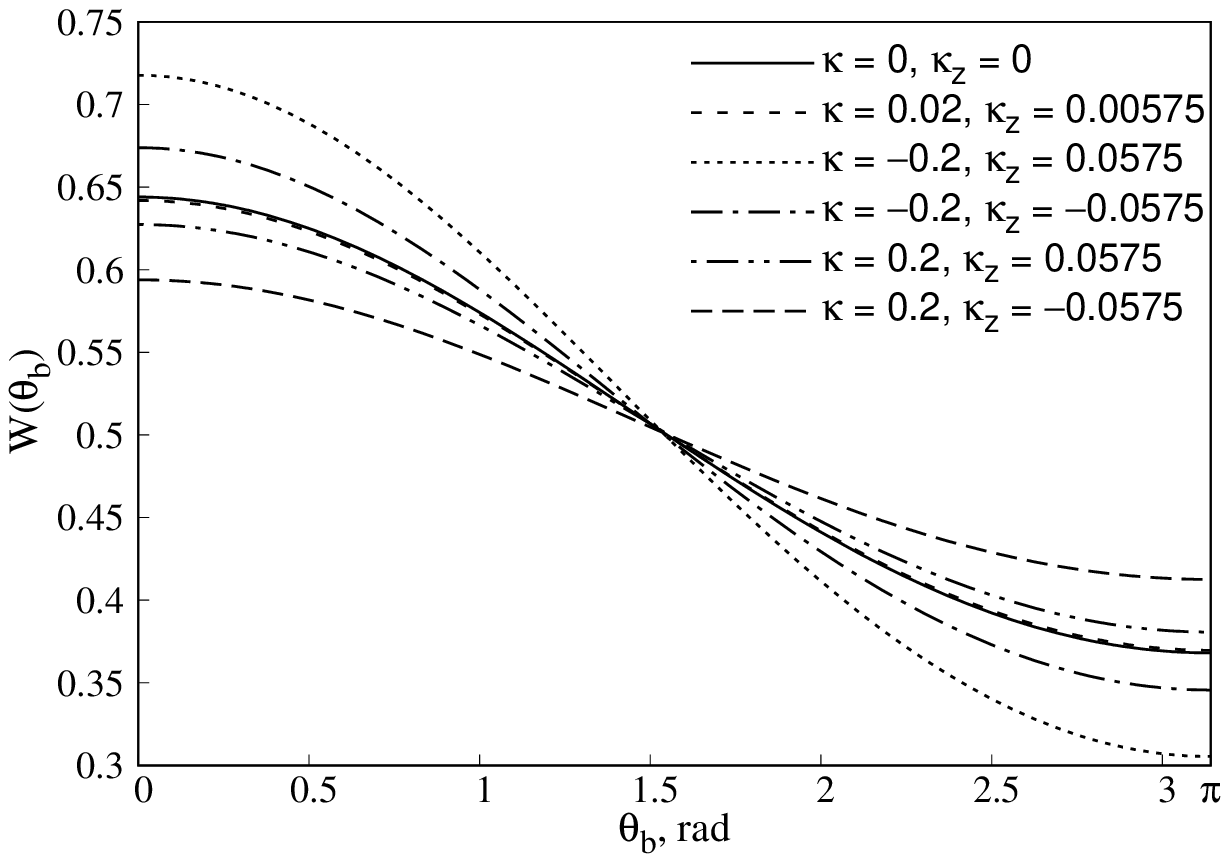}
}
\caption{Energy (left panel) and angular (right panel) normalized distributions of the bottom quark 
at various values of the couplings $\kappa$,  $\kappa_z$.  The $e^+e^-$ energy is 380 GeV. 
All curves are calculated for the polarized process.}
\label{fig:distributions}
\end{figure}

These distributions are shown in Fig.~\ref{fig:distributions}.
It is seen that for the energy distributions in Fig.~\ref{fig:distributions} (left), all the curves cross at the energy $\tilde{E}_b =\gamma p_b^0$. 
This behavior follows from Eq.~(\ref{eq:012}), in particular, the couplings $\kappa$ and $\kappa_z$ contribute only to 
the polarization-dependent part proportional to $E_b-\tilde{E}_b$ which vanishes at $E_b=\tilde{E}_b$.  The value of the 
distribution at the crossing point is $W(\tilde{E}_b)= (E_+ - E_-)^{-1} = (2 V \tilde{E}_b)^{-1}$, 
which is $ 0.0162$ GeV$^{-1}$ at the  $e^+e^-$ energy 380 GeV.

The angular distribution in  Fig.~\ref{fig:distributions} (right) has somewhat different features which follow 
from Eq.~(\ref{eq:019}).  The couplings $\kappa$ and $\kappa_z$ enter both polarization-independent and polarization-dependent parts of the distribution. 
All the curves cross at the angle $\tilde{\theta}_b \approx \pi/2$, and the value at the crossing point is $W(\tilde{\theta}_b)=1/2$ independently 
of the $e^+e^-$ energy.  It is also seen that the angular distribution is more sensitive to values of $\kappa$ and $\kappa_z$ than the energy distribution. 

These simple properties of the energy and angular normalized distributions make them convenient observables for future experimental studies.


\subsection{\label{subsec:asymmetries} Asymmetries}

Other important observables, sensitive to the BSM couplings, are the asymmetries of the cross-sections. 
In particular, for the cross-section in Eq.~(\ref{eq:012}) one can define the energy asymmetry 
\begin{equation}
 {\cal A}_{E} = \left. { \left(\int_{\tilde{E}_b}^{E_+} - \int_{E_-}^{\tilde{E}_b}  \right) dE_b \, \frac{d \sigma (e^+ e^- \to  b \, W^+ \, \bar{t} )}{d E_b} }
\right/ {\int_{E_-}^{E_+} dE_b  \frac{d \sigma (e^+ e^- \to  b \, W^+ \, \bar{t})}{d E_b}  }, 
\label{eq:021}
\end{equation}
where the energy $\tilde{E}_b = \gamma \, p_b^0$. This energy is determined by the $e^+ e^-$ energy $\sqrt{s}$.     

For the cross-section in Eq.~(\ref{eq:019})  the forward backward (FB) asymmetry can be defined as follows
\begin{equation}
 {\cal A}_{FB} =\left.    \left( \int_{0}^{+1}  - \int_{-1}^{0}  \right)  d z  \,
\frac{d \sigma (e^+ e^- \to  b \, W^+ \, \bar{t})}{d z }   \right/
 \int_{-1}^{+1}  d z  \, \frac{d \sigma (e^+ e^- \to  b \, W^+ \, \bar{t})}{d z }   
\label{eq:022}
\end{equation}  
with $z \equiv \cos \theta_b$.  Due to the normalizations (\ref{eq:014}) and (\ref{eq:020}) the denominators
of asymmetries (\ref{eq:021}) and (\ref{eq:022}) are fixed by the total $e^+ e^- \to t \, \bar{t}$ cross-section. 

The calculated asymmetries are presented in Table~\ref{tab:2}. The values of the BSM couplings are chosen as in 
Table~\ref{tab:1}. It is seen that for certain coupling constants the energy asymmetry reaches a few percent, and the 
angular asymmetry takes quite sizable values 10--20\% that could be accessible in future experiments.     

\begin{table}[tph]
\tbl{The asymmetries ${\cal A}_E$ and ${\cal A}_{FB}$ in \% for various values of the couplings $\kappa$ and $\kappa_z$ at the $e^+ e^-$ energy 380 GeV.}
{\begin{tabular*}{0.63\textwidth}{c@{\extracolsep{\fill}}c cc}
\toprule
$\kappa$ & $\kappa_z$ & ${\cal A}_E$  & ${\cal A}_{FB}$ \\
\colrule 
0.0   & 0.0  & 1.7 & 14.0 \\
0.02 & 0.00575 &   1.6  & 14.0  \\
0.2 & 0.0575 &  1.3  & 12.0 \\
0.2 & $-$0.0575 &  0.9  &  9.0 \\ 
$-$0.2 & 0.0575 &  4.0  & 21.0 \\ 
$-$0.2 & $-$0.0575 &  2.0  &  16.0  \\ 
\botrule
\end{tabular*} \label{tab:2}}
\end{table}

\begin{figure}[tbh]
\begin{center}
\includegraphics[width=0.70\textwidth]{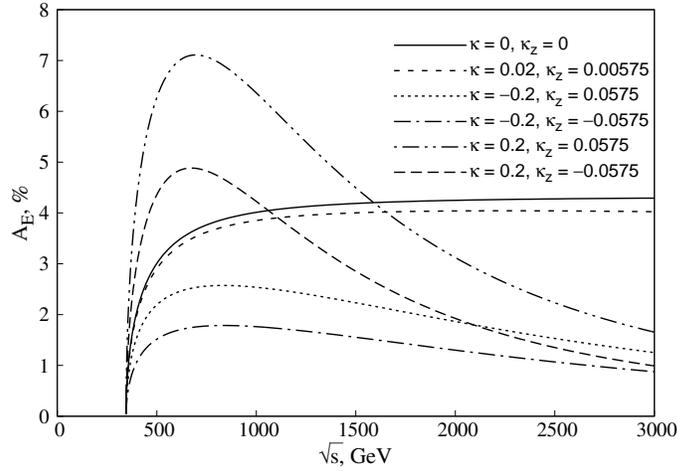}
\end{center}
\caption{The energy asymmetry as a function of $\sqrt{s}$  for various values of the  couplings.}
\label{fig:5}
\end{figure}

\begin{figure}[tbh]
\begin{center}
\includegraphics[width=0.70\textwidth]{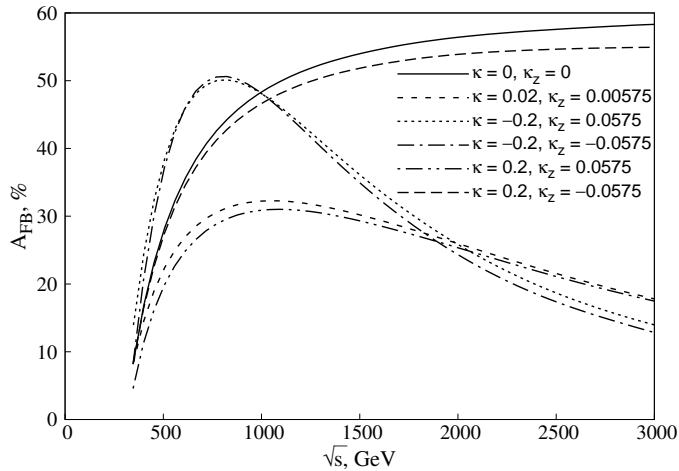}
\end{center}
\caption{The angular asymmetry as a function of $\sqrt{s} $ for various values of the couplings.  }
\label{fig:6}
\end{figure}

It is of interest to study dependence of the asymmetries on the $e^+e^-$ energy. The dependence of ${\cal A}_E$ on $\sqrt{s}$ 
is plotted in Fig.~\ref{fig:5}, and the corresponding dependence of  ${\cal A}_{FB}$ is plotted in Fig.~\ref{fig:6}.  

As follows from Fig.~\ref{fig:5}, the energy asymmetry in the SM rises up to the energy $\sim$ 1 TeV and then stays almost constant, 
while the asymmetry beyond the SM has another trend --  there is a wide maximum at the energy $\sqrt{s} \sim$ 650--850 GeV and then it decreases. 
This behavior can be of interest for the experimental studies at the CLIC, during the next stages of its run, 
in which the energy is planned to be 1.5 TeV (the 2nd construction stage) and 3 TeV (the 3rd construction stage) with 
the expected integrated luminosities of 2.5 ab$^{-1}$ and 5 ab$^{-1}$, respectively.     

Somewhat similar behavior is observed for the angular asymmetry in Fig.~\ref{fig:6}, though the values of ${\cal A}_{FB}$ are an order of magnitude larger than 
the values of ${\cal A}_E$. 

Finally, we briefly address the equivalence of the two methods of calculation of the cross-section mentioned in the 
beginning of Sec.~\ref{sec:formalism}. 
One has to check the equivalence of Eqs.~(\ref{eq:001}) and (\ref{eq:002}), though formally they look differently.     
To calculate the cross-section using Eq.~(\ref{eq:002}), we evaluate the $e^+ e^- \to t \, \bar{t}$ \ differential cross-section 
with the polarization density matrix of the top quark
\begin{equation}
u(p_t, \, m_t) \, \bar{u}(p_t, \, m_t) = \frac{1}{2}(\vslash{p}_t+m_t)  ({1 + \gamma^5 \vslash{n}})
\label{eq:023}
\end{equation}
with $\vslash{p}_t=\gamma \cdot p_t$, \ $\vslash{n} = \gamma \cdot n$ and the four-vector $n^\mu$ given in (\ref{eq:003}).  
Note that  in the top-quark rest frame  $n^\mu_{R}=(0, \, \alpha_b  \vec{n}_{b, R})$. 
      
Our explicit calculation of the differential cross-sections in Eq.~(\ref{eq:012}) and Eq.~(\ref{eq:019}) shows that equations 
(\ref{eq:001}) and (\ref{eq:002}) lead to the identical results, as expected. 


\section{\label{sec:conclusions} Conclusions}

We studied the spectra of the bottom quark from the decay of the top quark, $t \to b \, W^+$, produced in the 
electron--positron annihilation.  The formalism used in calculation is based on the method of Ref.~\citen{Kawasaki:1973hf}. 
The cross-sections of the process $e^+ e^- \to t \, \bar{t} \to  b \, W^+ \, \bar{t}$,  as functions of the $b$-quark energy and angle 
with respect to the direction of the electron beam, are derived and calculated.  In the calculation the $e^+ e^-$ energy 
$\sqrt{s}=380$ GeV corresponding to the first construction stage of the CLIC, was chosen.    

We investigated the influence of the top-quark polarization, which arises in the $e^+ e^- \to t \bar{t}$ reaction, on the energy and angular spectra 
of the bottom quark. In general, the difference between the cross-sections for the polarized and unpolarized top quark  
is not too big, the effect is of the order of 10\%.  
The angular spectrum turns out to be more sensitive to the top-quark polarization than the energy spectrum.  

It is shown that the cross-section of the $e^+ e^- \to  b \, W^+ \, \bar{t}$ reaction strongly depends on values of the $\gamma t \bar{t}$ and 
$Z t \bar{t}$  anomalous couplings $\kappa$ and $\kappa_z$. These couplings beyond the SM can take quite sizable values, 
and we studied how the energy and angular spectra of the $b$ quark depend on $\kappa$ and $\kappa_z$.
It follows from the calculations that the BSM effects can be quite important and perspective for studying the top quark properties. 

Several observables sensitive to these couplings were considered, namely, the energy and angular normalized distributions, 
and the energy and angular asymmetries. In particular, the angular  
asymmetry ${\cal A}_{FB}$ at $\sqrt{s}=380$ GeV reaches  10--20\%, that can possibly be accessible in future experiments. 
We also investigated dependence of these asymmetries on the invariant $e^+ e^-$ energy up to $\sqrt{s}=3$ TeV. 
An interesting  trend is observed --  for the couplings beyond the SM these asymmetries have a maximum 
at the energy $\sqrt{s} =$ 650--850 GeV and then they decrease, while in the SM the asymmetries 
slowly rise with $e^+ e^-$ energy and reach 
$\sim 4 \%$ for the energy asymmetry and $\sim 60 \%$ for the angular asymmetry.  
This behavior can be of interest for future studies at the CLIC at the next stages of its run, and for other $e^+ e^-$ colliders. 

Although the consideration in the paper was performed for the unpolarized electron and positron, the present method can easily be extended to the case 
of the polarized electron and positron beams. The next step in future can also be the study of the joint distributions of $b$ and $\bar{b}$ quarks  
from decays of the top quark and antiquark.


\section*{Acknowledgments}

This work was partially conducted in the scope of the IDEATE International Associated Laboratory (LIA).
The authors acknowledge partial support by the National Academy of Sciences of Ukraine 
via the program ``Support for the development of priority areas of scientific research'' (6541230).


\appendix

\section{Gauge Invariance of Matrix Element of the $e^+  e^- \to t  \, \bar{t} \to b \, W^+ \, \bar{t}$ Process \label{app:A}} 

In this Appendix we prove the gauge invariance of matrix element for the photon-exchange diagram in Fig.~\ref{fig:1} 
in the narrow-width approximation (NWA).  This approximation lies in the derivation of Eqs.~(\ref{eq:001}) and (\ref{eq:002})  
which are used in the calculation. 

This corresponding matrix element can be written as  

\begin{eqnarray}
{\cal M} &= &\frac{1}{q^2} \ell_\mu \, J^\mu, \quad     
\ell_\mu  =  e \, \bar{v}(k^\prime) \gamma_\mu u(k),  \qquad  \quad  \nonumber \\
J^\mu &=& \frac{eg}{2 \sqrt{2}} \frac{1}{p_t^2 -m_t^2 + i m_t \Gamma_t} \, I^\mu,  \label{eq:A1} \\
I^\mu & = & \bar{u} (p_b) \, \vslash{\epsilon}^* (p_W)  (1- \gamma_5) (\vslash{p}_t + m_t) 
V_{\gamma t \bar{t}}^\mu \, v(p_{\bar{t}})  \nonumber
\end{eqnarray} 
with the energy momentum conservation $k+k^\prime \equiv q = p_t + p_{\bar{t}} = p_b + p_W +  p_{\bar{t}}$,  \ 
$\vslash{\epsilon}^* (p_W) = \gamma^\alpha \epsilon_{\alpha}^* (p_W)$, where $\epsilon (p_W)$ is the polarization four-vector of the $W$ boson. 
The spinors of the electron and positron are $u(k)$  and $v(k^\prime)$, and those of the bottom quark and top antiquark are $u (p_b)$ and $v(p_{\bar{t}})$.    
In Eq.~\ref{eq:A1} the propagator of the intermediate top quark  is taken in the form 
\begin{equation}
S(p_t) = \frac{\vslash{p}_t + m_t }{p_t^2 -m_t^2 + i m_t \Gamma_t}
\label{eq:A2}
\end{equation}
with the  top quark total decay width $\Gamma_t$. 

The $\gamma t \bar{t}$  vertex $V_{\gamma t \bar{t}}^\mu$ for the on-mass-shell quarks can be chosen in the general form~\cite{Hollik:1998vz}
\begin{equation}
V_{\gamma t \bar{t}}^\mu = F_1(q^2) \gamma^\mu + i \frac{\sigma^{\mu \alpha} q_\alpha}{2 m_t} [F_2 (q^2) - i \gamma_5 F_3(q^2)].  
\label{eq:A3}
\end{equation}
 With the proper normalization of the form factors $F_{1,2,3}(q^2)$ at $q^2=0$, Eq.~(\ref{eq:A3})  reduces to the vertex in Eq.~(\ref{eq:020_vertex_g}).

After squaring the matrix element we get 
\begin{eqnarray}   
|{\cal M}|^2 = \frac{e^4 g^2}{8 q^4 \, [(p_t^2 -m_t^2)^2 + m_t^2 \Gamma_t^2] } \, \ell_\mu \, \ell_{\nu}^* \, W^{\mu \nu},    \quad 
W^{\mu \nu} = I^\mu I^{\nu *}.
\label{eq:A4}
\end{eqnarray}

In the NWA one applies the relation (see,  for example Refs.~\citen{Kawasaki:1973hf, Arens:1994jp})
\begin{equation}
\frac{1}{(p_t^2 -m_t^2)^2 + m_t^2 \Gamma_t^2} \approx \frac{\pi}{m_t \Gamma_t} \, \delta(p_t^2 -m_t^2),
\label{eq:A5}
\end{equation}
valid for $\Gamma_t \ll m_t$. Note that accuracy of the NWA has been studied in various 
processes~\cite{Achasov:1995ac, Berdine:2007uv, Uhlemann:2008pm, Zagoskin:2015sca}.

Let us now check the gauge invariance. We use the condition $q_\mu V^\mu_{\gamma t \bar{t}}  = F_1(q^2) \vslash{q} 
= F_1(q^2) (\vslash{p}_t + \vslash{p}_{\bar{t}}) $ 
and obtain
\begin{equation}
q_\mu W^{\mu \nu}= F_1(q^2)  (p_t^2 - m_t^2) \,  \bar{u} (p_b) \, \vslash{\epsilon}^* (p_W)  (1- \gamma_5) v(p_{\bar{t}}) \, I^{\nu *},
\label{eq:A6}
\end{equation}
and therefore in view of (\ref{eq:A4}) and (\ref{eq:A5}), $q_\mu W^{\mu \nu} =0$. Clearly, $q_\nu W^{\mu \nu} =0$ as well.

We emphasize that this result is a consequence of the NWA which implies that the intermediate top quark is on mass shell, $p_t^2 = m_t^2$. 
Otherwise the gauge invariance would be violated, and additional tree-level diagrams discussed in Refs.~\citen{Boos:1999qc, Boos:2001sj, Boos:2012vm} 
would be needed to restore the gauge invariance. 

It is interesting to note, that in the present approach the gauge invariance holds for arbitrary form factors $F_{1,2,3}(q^2)$ which can include 
radiative corrections and effects beyond the SM.


\bibliographystyle{ws-ijmpa}
\bibliography{bib}
\label{17}
\end{document}